# Chapter 8

# Tactics for Internal Compliance: A Literature Review

Ralph Foorthuis

*Compliance of organizations with internal and external norms is a highly relevant topic for both practitioners and academics nowadays. However, the substantive, elementary compliance tactics that organizations can use for achieving internal compliance have been described in a fragmented manner and in the literatures of distinct academic disciplines. Using a multidisciplinary structured literature review of 134 publications, this study offers three contributions. First, we present a typology of 45 compliance tactics, which constitutes a comprehensive and rich overview of elementary ways for bringing the organization into compliance. Secondly, we provide an overview of fundamental concepts in the theory of compliance, which forms the basis for the framework we developed for positioning compliance tactics and for analyzing or developing compliance strategies. Thirdly, we present insights for moving from compliance tactics to compliance strategies. In the process, and using the multidisciplinary literature review to take a bird's-eye view, we demonstrate that compliance strategies need to be regarded as a richer concept than perceived hitherto. We also show that opportunities for innovation exist.[1]*

## 8.1 Introduction

With the advent of stricter legal demands, industrial best practices, security concerns, ethical codes of conduct and IT management standards, the topic of internal compliance has become highly relevant for both practitioners and academics (MacLean and Behnam, 2010; Short and Toffel, 2010; OCEG, 2009; Cleven and Winter, 2009; Schneiberg and Bartley, 2008; Harris and Cummings, 2007; Tyler and Blader, 2005; Emmerich et al., 1999). Simply put, internal compliance aims at ensuring that organizational actors' behaviors and outputs conform to the norms (see also 8.3.1). This form of compliance can relate to various types and levels of prescriptive systems that guide and constrain organizations. International and domestic laws and regulations, industry-wide standards and best practices, and organizational guidelines, rules and procedures all require organizational actors to conform to norms. Different types of prescriptive systems have different raisons d'être. Laws and regulations are mostly concerned with mitigating societal risk and encouraging ethical behavior (MacLean and Behnam, 2010; Sadiq and Indulska, 2008). Industrial best practices and organizational procedures usually aim at mitigating organizational risk, improving competence, effectiveness and efficiency, and achieving compliance as an end in itself (Szulanski, 1996; O'Dell and Grayson, 1998; COSO, 1994). Business and IT management standards, such as Enterprise Architecture principles, mainly focus on achieving organization-wide business goals, alignment, integration, complexity control and agility (Lankhorst et al., 2005; Gregor et al., 2007; Boh and Yellin, 2007).[2]

---

1 A preliminary version of the framework has been published as: Foorthuis, R.M., Bos, R. (2011). *A Framework for Organizational Compliance Management Tactics*. In: GRCIS 2011 (CAiSE 2011 Workshop), LNBIP 83, pp. 259–268. Berlin: Springer-Verlag. The typology of tactics and the guidelines for developing compliance strategies have been submitted to a journal.

    The topic of compliance has fascinated scholars for centuries. As early as the 1600s, Thomas Hobbes touched on the delicate issue of the *compliance problem* (Gauthier, 1991; Hollis, 1994; Hartman, 1996). He stated that, although compliance with 'contracts' may be better for the group as a whole and it may be in an individual actor's best interest to agree to contracts, it may very well not be in his interest to actually comply with them. Following this logic, it is necessary for internal regulators to actively pursue and assess compliance. This is also true in an organizational context, as compliance with the norms may be in the best interest of the organization as a whole, but may not lead to optimal results from the perspective of complying individuals, projects or departments. This is not merely a philosophical stance, given the scandals society has seen at organizations such as Enron, WorldCom and Parmalat (Kump and Rose, 2004; El Kharbili et al., 2008; MacLean and Behnam, 2010). Moreover, several studies demonstrate that non-compliance in organizations is widespread (Meyer and Rowan, 1977; Healy and Iles, 2002; Malloy, 2003; Tyler and Blader, 2005). This makes compliance a strategic issue in the current era, especially considering the high financial and non-financial costs organizations have to pay for their

---

2  Note that, although the previous chapters of this doctoral thesis focus on compliance with Enterprise Architecture, this chapter investigates compliance in general.

non-compliance. With regulations such as the Sarbanes-Oxley Act, transgressing organizations and individual CEOs and CIOs face severe penalties, such as fines and imprisonment (Volonino et al., 2004; Braganza and Franken, 2007). Scandals and unethical firm behavior can also severely damage an organization's reputation due to dissatisfied customers, shareholders, employees and other stakeholders (Philippe and Durand, 2011; Rossouw and Van Vuuren, 2003; Harris and Cummings, 2007). Demonstrating compliance with regulations, industrial best practices and ethical norms, on the other hand, can yield not only legitimacy, but also a good reputation and the benefits that come with that, such as attracting large institutional investors and customers (ibid., Meyer and Rowan, 1977; Emmerich et al, 1999; Malloy, 2003; Currie, 2008; El Kharbili et al., 2008).

Organizations experience difficulties, however, in implementing their compliance management approaches (Hurley, 2004; Sadiq and Indulska, 2008; MacLean and Behnam, 2010), possibly due to a lack of awareness of the full spectrum of actions that can be taken (cf. Straub and Welke, 1998). At the same time, both the compliance stimulating tactics (i.e. compliance techniques such as training and penalties) that comprise these approaches and the concept of compliance itself have been described in the extant literature in a fragmented manner and from very different perspectives. Consequently, there is the need for a structured overview of generic ways for encouraging compliance (cf. Cleven and Winter, 2009). This chapter therefore aims to answer the following research question:

> *What compliance tactics can be used by an organization to increase, achieve or maintain compliance with internal and external norms?*

A structured overview of such techniques, aiming to be as complete as possible and drawing upon multiple disciplines, has to the best of our knowledge not been published before. We will present the inventory of 45 generic compliance tactics as a typology, based on a multidisciplinary literature review of 134 publications. This chapter will also deliver two additional supporting contributions. First, we present an *overview of fundamental insights into compliance*. Along with definitions of key concepts, this will provide the basis for a conceptual structure or *framework* that can be used as an analytical tool for classifying, studying and presenting existing compliance-stimulating tactics. This structure will therefore be used for creating the typology of compliance tactics. In a process of 'picking and choosing' from the typology, such tactics form the fundamental elements of a broader compliance strategy. Therefore, as a second additional contribution, we will present a discussion on how internal regulators can use the typology to develop an overall and coherent organizational *compliance management strategy*. In addition to this practical purpose underlying our study, the classic and state-of-the-art insights presented in this chapter serve as an introduction to the topic of compliance for newcomers. Academics and practitioners already acquainted with the field may find that this chapter broadens their horizons.

It is important to note that the focus of this study is on *internal compliance*, i.e. organizational compliance. The organization is not a black-box in this study. Our perspective is not that of governments and other societal institutions, but that of an organization's internal regulators – such as senior managers, policy drafters and compliance officers – aiming at a compliant organization. The organization-wide policies and norms are created and implemented by these regulators. However, the organization does not operate in a vacuum and named creation and implementation processes are driven largely by external pressures from governments, competitors, customers, suppliers, partners and institutions such as NGOs and labor unions (cf. DiMaggio and Powell, 1983; Scott, 2003; Meyer and Rowan, 1977; Currie, 2008; Short and Toffel, 2010; Weaver, Trevino and Cochran, 1999a; Ward and Peppard, 2002). Figure 8.1 shows the organization, its internal actors and typical pressures exerted by external stakeholders.

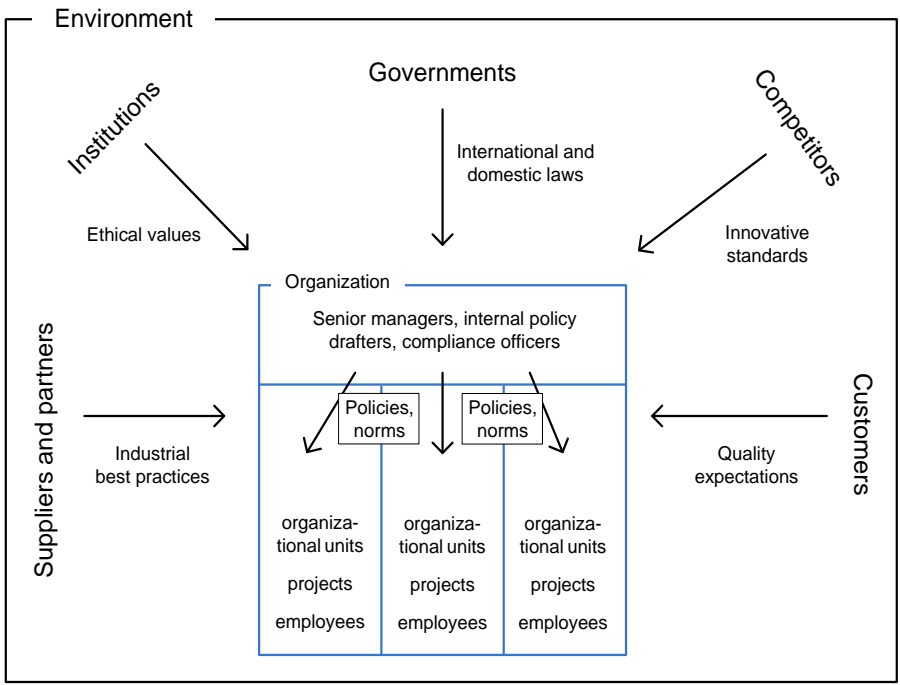

Figure 8.1. The organization and its normative environment

After determining and interpreting the relevant pressures and translating them into internal policies, the various organizational units, projects, programs – and the individuals of which they are comprised – are expected to conform to them. Although organizations have been known to adopt symbolic compliance measures instead of substantively changing their operations (MacLean and Behnam, 2010; Short and Stoffel, 2010; Meyer and Rowan, 1977), we focus here on organizations that genuinely strive to be compliant. It is here that the underlying goal of this chapter lies, as we investigate what tactics an organization has at its disposal for achieving a satisfactory level of compliance.

A concept commonly used when discussing compliance measures taken by an organization is that of internal controls. An *internal control* is usually defined as a formal process designed to provide reasonable assurance of achieving effective and efficient operations, reliable (financial) reporting, and compliance with laws and regulations (COSO, 1994; Luthy and Forcht, 2006; Namiri and Stojanovic, 2007). However, in this study we will introduce the concept of *compliance tactic* instead, which we define as a measure that can be taken, or a technique or mechanism that can be used, to encourage compliance of relevant organizational actors. The concept of tactics allows us to identify the theoretical essence of compliance endeavors rather than their practical implementation. See section 8.3 for more on this.

This chapter proceeds as follows. In section 8.2, our research approach is described. Section 8.3 defines and discusses fundamental compliance concepts. Section 8.4 introduces the conceptual structure and typology. Section 8.5 presents a detailed description of the typology's compliance tactics. Section 8.6 discusses findings of the literature review and presents insights for developing a compliance management strategy. Section 8.7 is for conclusions.

## 8.2  Research approach

We employed a structured literature study for our research, as this provides an appropriate method to investigate the fundamentals of compliance and enables us to identify the wide array of techniques devised in distinct disciplines. Science can benefit from drawing from different fields, as a topic can be enriched by the exposure to distinct and potentially relevant theoretical backgrounds (cf. Webster and Watson, 2002; Malone and Crowston, 1994). The disciplines we have drawn upon are those of *management* and *business studies* (e.g. corporate compliance and ethics programs), *law* and *political science* (e.g. environmental agreements and determinants of compliance), *information systems* (e.g. security policies and informational privacy), *philosophy* (e.g. business and contractarian ethics), *organizational sociology* (e.g. institutional factors), *economics* and *accounting* (e.g. compliance audits), *engineering* (e.g. technology for data protection), and *social psychology* (e.g. influence mechanisms).

Based on Tranfield et al. (2003) and Brereton et al. (2007), we have structured our literature study as follows. In the first phase, the research project was planned. This entailed specifying initial research questions, gathering knowledge on methodological aspects of literature reviews, and specifying search criteria for identifying relevant publications. Since the identification of compliance tactics would be a crucial part of the study, we concluded that publications should be searched for in several academic disciplines. In addition, as the nature of the study was less a 'truth finding' mission than a broad and open-minded identification effort, quality criteria for journals and conferences could not be too strict (i.e. limiting the reviewed publications to only a handful of the fields' top journals was not an option). Nonetheless, the norm was that a publication be academically peer reviewed, unless a technical report or practitioner publication yielded a significant insight.

In the second phase, potentially relevant literature was identified and collected. Key search terms were "compliance", "conformance" and "conformity". Other terms used were "compliance strategy", "compliance management", "compliance policy", "compliance auditing", "corporate compliance" and "organi[z|s]ational compliance". The search was conducted in academic indexing services, such as JSTOR, EconPapers and PiCarta, in broader listings such as GoogleScholar, and in search engines of specific journals. Title, abstract, keywords and occasionally conclusions informed the decision of whether or not to include the paper in the set of candidate publications. The literature was collected by the principal researcher and an information specialist of Statistics Netherlands' library. All in all, over 200 (mostly digital) publications were collected in this phase.

In the third phase, a literature database was created for systematic storage of publication characteristics to be gathered during the actual review process undertaken in the fourth phase. The database was based on the concept matrix of Webster and Watson (2002) and data extraction guidelines of Tranfield et al. (2003) and Kitchenham et al. (2009). It allowed storage of information such as the publication's title, authors, unit of analysis and substantive conceptual contribution. In addition, a review protocol was established to ensure that the review process would be carried out in a systematic fashion (see Appendix 8.A). The protocol contained the inclusion and exclusion criteria for the definitive selection of publications, the steps of the review process, additional conditions, definitions and data extraction guidelines (cf. Brereton et al., 2007). In addition, several publications were studied or roughly scanned in this phase, both to obtain fundamental insights and to identify promising additional references. A preliminary version of the analytical structure was created, as this would function as the two-dimensional framework in which the tactics would be positioned. In order to prevent leaving the reader with only a set of fragmented tactics, the research goal was also complemented with the sub-goal regarding the development of compliance strategies.

In the fourth phase, the actual reviewing of the collected publications was conducted, i.e. identifying the compliance tactics and gathering insights into compliance strategies. Each decision to review a specific publication was based on the likelihood that it provided a new tactic or deeper insights into strategies or known tactics. A total of 134 publications was reviewed (excluding methodological and peripheral articles). See Tables 8.1 to 8.4 for an overview. The reviewing process was conducted by the principal author. Journal articles and conference papers were reviewed fully, whereas of books only the relevant sections were analyzed. After reviewing a publication, a descriptive record was added to the literature database (see Appendix 8.B for several examples). The identified compliance tactics, if any, were paraphrased and added to the preliminary tactics overview. Subsequently, the text in this document was iteratively coded to develop the typology of tactics (see section 8.4 for a definition of typologies). We used the method of Miles and Huberman (1994) for this coding process, as it aims at creating a classification or typology (in contrast to Grounded Theory, in which codes are developed to create a causal theory). A code in our study represented a (candidate) tactic. Two kinds of coding processes were used in the study. First-level coding, working mainly on the source text, is the process of creating summarized pieces of data and their respective codes. The first-level coding activities resulted in the creation of a preliminary classification of tactics. After 35 publications had been reviewed and coded in this manner, an iterative and creative process of pattern coding was initialized to run parallel with the continuing review and first-level coding activities, resulting in a more mature set of tactics. Pattern codes are descriptive pieces of data and their respective codes, representing patterns found in the underlying publications. Based on the first-codes, they evolved into the final versions of the compliance tactics presented in section 8.5. In the process, first-level codes could be split up into several pattern codes, or they could be combined into a single pattern code (see Appendix 8.C for an overview of splits and combinations). This was due in part to the fact that the first-level codes were based more directly on the underlying individual publications, whereas the pattern coding was aimed at generalizing the tactics across all publications. Part of the latter process was also to position the tactics within the analytical structure (the two-dimensional framework presented in section 8.4). Important criteria for defining a definitive (pattern code) tactic were mutual exclusiveness and a conceptual abstraction that allowed for a clear and presentable descrip-tion. The framework was also finalized during this phase, after going through several iterations. Furthermore, knowledge on compliance management and strategies (presented in section 8.6) was gathered in this phase, as part of the review process.

The fifth phase was writing the review report, which culminated in this chapter. An initial and short version was presented as a paper at an international workshop on compliance in order to obtain feedback.

| Discipline | Number of publications |
|---|---|
| Management and Business | 34 |
| Law and Political Science | 19 |
| Information Systems | 30 |
| Philosophy and Sociology | 22 |
| Engineering | 15 |
| Social Psychology | 6 |
| Economics and Accounting | 8 |
| Total | 134 |

Table 8.1. Number of publications per discipline

| Period | Number of publications |
|---|---|
| 1970s and before | 3 |
| 1980s | 5 |
| 1990s | 34 |
| 2000s | 79 |
| 2010s | 13 |
| Total | 134 |

Table 8.2. Number of publications per period

| Concept | Number of publications |
|---|---|
| Fundamental compliance concepts (sections 8.3 and 8.4) | 63 |
| Compliance tactics (section 8.5) | 91 |
| Compliance strategy (section 8.6) | 52 |

Table 8.3. Number of publications per concept

| Level of analysis | Number of publications |
|---|---|
| Individual employee | 44 |
| Collective (projects, organizational units) | 60 |
| Organization | 87 |
| (Inter)national bodies | 24 |

Table 8.4. Number of publications per analytical level

## 8.3 Fundamentals of Compliance

This section defines and discusses key concepts that function as the basis for our study, such as compliance, conformity, actors, norms and policies. The nature of compliance is also explored, resulting in the insights used to structure the two-dimensional framework that forms the foundation for our typology.

### 8.3.1 Compliance – Definitions and Key Concepts

There is no generally accepted definition of the term *compliance* (Faure and Lefevere, 2005; Pupke, 2008). Many authors define compliance as (the extent to which there is) adherence to laws and regulations (e.g. Zaelke et al., 2005; Faure and Lefevere, 2005; Mitchell, 1996; Schneiberg and Bartley, 2008) or to prescriptions in general, including industrial standards and internal policies (e.g. Kim, 2007; Daniel et al., 2009). Some authors also seem to include unwritten, ethical norms (e.g. Welcomer, 2002; Badea and Pana, 2010). Other authors go beyond compliance as a status-oriented concept and define the term as a process (e.g. Caldwell and Eid, 2007; Meeuwisse, 2010) or ability (e.g. IIA, 1997), or emphasize in their definitions the organizational structures and processes to ensure adherence to them (cf. Pupke, 2008). The term compliance has also been extended to include underlying goals, such as transparency and protecting an organization's reputation (e.g. Panitz et al., 2010; Jourdan and Oehler, 2005; Pupke, 2008).

In the literature there may be even less agreement on the term *conformity* (cf. Bond and Smith, 1996). Similar to compliance, conformity is regularly used as adherence to prescribed rules (Merton, 1957; Schapiro, 2003; Currie, 2008; Mitchell, 1996; Ellis et al., 1993; Tyler and Blader, 2005). Conformity has also been contrasted with compliance, with the latter following an explicit or implicit request, and the former referring to a state of accordance in the absence of a request (Cialdini and Goldstein, 2004). In this context, conformity is sometimes said to necessarily involve a change in belief or behavior (Zimbardo and Leippe, 1991; Rowe, 2005), whereas an actor can be compliant without a position change (see below). In addition, conformity is seen as a broad concept that allows both public agreement and internalization of norms, whereas compliance represents public agreement without private agreement (Levine and Resnick, 1993; Bond and Smith, 1996). In addition, conformity is often used to specifically refer to the phenomenon that organizations within a particular institutional sphere tend to become more similar as time progresses (Scott, 2003; DiMaggio and Powell, 1983). Finally, conformity is also seen as being comprised of both procedures and underlying goals (Philippe and Durand, 2011). The term *conformance*, likewise, is not used in a single, specific manner. For example, Falkl (2005) distinguishes between compliance and conformance, with the former referring to a binary adherence to a strict norm (necessitating a pass-or-fail outcome) and the latter describing how well the norms are matched (allowing for adherence of certain aspects and non-adherence of others). Chopra and Singh (2006) use conformance to refer to a (software) agent adhering to the norms in terms of its design, whereas compliance refers to this issue in terms of an agent's behavior. Other publications also use the term conformance in different ways (cf. Merton, 1957; The Open Group, 2009; Alter and Wright, 2010; Currie, 2008; Panitz et al., 2010). Given the lack of agreement illustrated above, we will not distinguish between the three terms in our study unless explicitly specified.

In this study we define *compliance* as the extent to which there exists a state of accordance between an actor's behavior or products on the one side and predefined, relevant and explicit norms on the other (cf. Zaelke et al., 2005; Kim, 2007; Faure and Lefevere, 2005; Mitchell, 1996; Abdullah et al., 2009). Norms in this respect can be rules, procedures, conventions, standards, guidelines, principles, legislation or other prescriptions. Although we do not focus on compliance with the implicit, broader spirit of the norms, we do acknowledge (relatively high-level) principles as norms – on the condition that they be made explicit. Unwritten ethical norms are therefore also excluded from our definition. Although such norms can be external input from the organizational environment, in the context of this study they should be explicitly specified in an internal policy. A compliant state can be achieved regardless of the motivations, causes or circumstances that have led to it (Zaelke et al., 2005; Mitchell, 1996). In our view, therefore, an actor can be compliant without internalizing the norms and without necessarily changing his beliefs or behavior. Furthermore, compliance includes adherence that is unintentional and of which the regulated actor is unaware. Finally, compliance should be distinguished from effectiveness, as a compliant state need not necessarily result in achieving the desired end goals (Mitchell, 1996; Faure and Lefevere, 2005; Zaelke et al., 2005). To achieve compliance, an organization usually requires active *compliance management*, which we define as the organizational processes and mechanisms intended to avoid, provide insight in, and deal with violations of and encourage compliance with relevant internal and external norms (cf. Abdullah et al., 2009; Caldwell and Eid, 2007). To prevent unethical business conduct and avoid regulatory penalties and loss of reputation, compliance management includes implementing structures and processes (see section 8.6.2). However, there are many more measures that can be taken. It is the primary aim of this chapter to identify these compliance tactics.

The term *norms* (or *prescriptions*) serves as a general denotation that encompasses more specific forms such as laws, standards, rules, principles and guidelines. Therefore, norms can refer to general, abstract (but explicit) principles or to detailed rules – or anything in between. They can also refer to prohibitive norms (so-called proscriptions). Norms can thus be both prescribing and constraining by nature. Furthermore, they can be legally required or voluntary by nature. Norms can relate to both behavior and products. Requiring a project to use the organization's standard system development method is an example of norms relating to behavior. Requiring the products produced by a business process to comply with organization-wide quality standards is an example of norms relating to products. Finally, prescriptions can (and probably will) change as time progresses. A coherent and goal-oriented set of norms is referred to here as a *policy* (cf. Garner, 2004; Kagal et al., 2003; ILRI, 1995).

When applying norms or assessing them on conformance, several aspects or dimensions should be taken into account (Foorthuis et al., 2012). A prescription should be applied *correctly*. Its use, or lack of it, should also be *justified* in the respective situation. Another issue is whether related prescriptions are applied *consistently*. A final concern is whether the *complete* set of (mandatory) norms is applied, as opposed to merely a convenient subset. Given the latter, one relevant issue concerns whether norms are mandatory or not. In practice, quite some prescriptions prove to be voluntary in nature, e.g. industrial best practices. Adherence to the norms then is more akin to the narrow sense of conformity as defined by Cialdini and Goldstein (2004), i.e. adherence without a request. In other cases, such as internal guidelines, there may be a request, but the norms may have a comply-or-explain status. Even when norms are truly mandatory, they are not always perceived as such in practice (Boss et al. 2009).

An *internal regulator* is a policy maker and upholder, such as a senior manager, policy drafter or compliance officer. We define an *actor* as an organizational entity acting within an organization, and equipped with cognitive capabilities, preferences, beliefs, values and at least some degree of autonomy (Hollis, 1994; Jones, 1991; Thiroux, 1977). As such, an actor can be e.g. an organizational unit, a project or an individual employee. In the context of this chapter, an actor is expected to comply with the norms (i.e. he is a regulated actor or regulatee). See section 8.3.3 for more on actors.

A distinction can be made between two types of non-compliance (cf. Schapiro, 2003; Merriam-Webster, 2011). First, a *transgression* (or *violation*) refers to a situation in which a norm is not complied with, e.g. by breaking a law or rule. A reason for this might be that the actor in question had no interest in conforming to this specific norm or simply did not know how to comply. Secondly, *subversion* refers to a situation in which an actor, for his own individual interest, attempts to undermine the entire compliance system itself, or at least an essential part of its norms. For example, when the implementation of standards is carried out in such a fashion as to demonstrate their inferiority and the need to abandon them altogether. In an organizational context, subversion might point to fundamental political problems, structural conflicts of interest or competing norm systems. An organization that is the result of a merger, for example, may have competing sets of architectural standards.

We define a *compliance tactic* as a measure that can be taken, or a technique or mechanism that can be used, to encourage compliance of relevant organizational actors. Tactics can range from top management's organization-wide governance mechanisms to lower management localized measures. Furthermore, a tactic can be preventative, detective and corrective in nature (Sadiq and Indulska, 2008; Straub and Welke, 1998). Tactics can also be formal or informal by nature, and they need not be processes. As one tactic is typically not sufficient to obtain compliance, multiple tactics need to be combined into a coherent strategy. A *compliance management strategy*, therefore, is a general plan or pattern featuring a coherent set of compliance tactics that aims to bring the organization to a state that is compliant with relevant norms, at least to a sufficient level (cf. Quinn, 1996; cf. Mintzberg, 1987; see section 8.6 for an elaborate discussion on compliance strategies). Such a strategy may aim for *holistic compliance*, which incorporates three elements: coherent instead of fragmented compliance efforts (Volonino et al., 2004; Cleven and Winter, 2009), a long-term scope (Sadiq and Indulska, 2008), and the ability of the approach to cover multiple laws, standards frameworks or internal procedures simultaneously.

In section 8.1 we have stated that we will use the concept of *compliance tactics* instead of *internal controls*. There are several reasons for this. First and foremost, our use of tactics enables us to focus on the substantive essence of compliance endeavors, whereas internal controls focus attention to practical implementation. Second, we focus on identifying ways for stimulating compliance, i.e. more or less generic and atomic elements that can be included in a broader compliance management strategy. An internal control, however, is not such a pure and elementary measure, but an aggregate concept in itself. Per definition, it comprises multiple elements that should be considered as separate tactics in a list of re-usable measures (such as monitoring and ensuring employee competency). A third reason why we focus on compliance tactics instead of internal controls, is that the latter concept is too narrow as a result of focusing on formal processes, thereby excluding several relevant compliance measures. Tactics, on the contrary, can be formal or informal, and they are not required to be processes. A fourth and final reason is that, in another sense, the concept of internal controls is too broad for the purpose of this study, focusing not only on compliance but also on other goals, such as general risk mitigation and efficiency and effectiveness of business activities.

### 8.3.2 The Nature of Compliance

In this section we will discuss some fundamental insights into compliance that will form the basis for our two-dimensional framework. In different fields, the literature on compliance distinguishes between two broad types of theory, namely rationalist and normative approaches (Chayes and Chayes, 1993; Grossman and Zaelke, 2005; March and Olsen, 1998; Malloy, 2003; Hollis, 1994; Li et al., 2010; Bulgurcu et al., 2010; Tyler and Blader, 2005; Zaelke et al., 2005; Mitchell, 1996). These theories provide distinct insights into compliance-related behavior and underlying motivations of states, organizations and individuals. The *rationalist view* focuses on the actor's calculation of benefits and costs in his decision on whether or not to comply. This approach uses a "logic of consequences", which sees actors as choosing rationally among alternatives. Game theory is a regularly used lens here to analyze behavioral motivations, using the prisoner's dilemma to model the Hobbesian compliance problem described in this chapter's Introduction (Kraus and Coleman, 1987; Hollis, 1994). In a rationalistic perspective incentives and disincentives will alter the outcome of the actor's calculation. Therefore, one major approach used here is enforcement (or command-and-control), in which *punishment* is used to deter unwanted behavior. *Rewards* are an additional means in the rationalistic perspective, increasing compliance by changing the cost-benefit calculation to the actor's advantage.

As a second perspective, the *normative view* focuses on cooperation and assistance as a way of encouraging compliance (Mitchell, 1996; Zaelke et al., 2005; Malloy, 2003; Chayes and Chayes, 1993). This approach uses a "logic of appropriateness", which views actions as based on identities, roles, obligations, and considerations of appropriate, fair and legitimate action. Normative theories do not take the stance that an actor's behavior is irrational, but tend to broaden the scope to prevent reducing the discussion to costs and benefits. Actors are seen as following the institutionalized rules and practices that link particular identities to particular situations. These rules need to be internalized and viewed as legitimate by those subject to them. Not restricted by the somewhat cynical

Hobbesian perspective, it is acknowledged that compliance may be hindered if rules are ambiguous, complex or continuously changing, or if they are too numerous or not easily available. Non-compliance may also be the inadvertent result of deficient routines or a lack of capacity, knowledge or commitment. For all these reasons, non-compliance should be *managed* instead of being sanctioned. Methods to increase compliance therefore often focus on increasing the actor's capacity to comply. This is effectuated by cooperating, providing support and encouraging shared discourse in order to render rules clearer, more persuasive and easier to commit to.

Rational and normative models are not mutually exclusive, but rather complement each other and provide different lenses for analyzing influences on compliance behavior (Tyler and Blader, 2005; Welcomer, 2002; Zaelke et al., 2005; March and Olsen, 1998; Malloy, 2003). Both perspectives are relevant to our research. For example, organization-wide standards may be dismissed for rational reasons, as conforming to them may take additional time and effort. Or it may be that the organizational units and employees conform to industrial standards because they value their identity as "professionals" or their role as "managers". Governmental organizations may also feel "obliged", or consider it "appropriate" not to spend tax payers' money unnecessarily. Both perspectives should therefore be taken into account.

### 8.3.3    Levels of analysis

Although theories from both the rational and normative perspective often regard actors that need to comply as unitary agents, a comprehensive perspective on compliance also needs to be able to disaggregate an actor into multiple sub-actors (Chayes and Chayes, 1993; Malloy, 2003; Zaelke et al, 2005). An organization is comprised of structural units, such as divisions and their sub-units, and of temporary initiatives, such as programs and their projects and teams. Furthermore, all of these entities will have individual members. Motivations for compliance-related behavior may differ between these different sub-actors (Malloy, 2003; Braganza and Franken, 2007), thus requiring different compliance measures. We deal with this issue in our study by acknowledging three conceptual levels. First, the level of the enterprise as a whole, in which "enterprise" can be taken to mean the entire organization, a division or even a network of organizations (cf. The Open Group, 2009). This is the level at which the internal regulators are located and at which the policies are determined – although there may obviously be pressure from higher (external) levels, which are out of scope here. The second level accommodates various types of collectives that are expected to comply. They exist within the enterprise, such as departments and their sub-units and programs and their projects and teams.

These collectives typically have a more local scope and may have a political agenda that can, at least in part, be inconsistent with the wider enterprise and its policies. The third level is that of individuals, who may be directly expected to comply (e.g. in the case of information security procedures) or who may be part of a collective that is requested to comply (e.g. in the case of a project implementing a medical administration system that needs to comply with privacy regulations). In both cases, the decisions and behavior of individuals are determinants of actual compliance. However, although an individual may be inclined to comply, the project of which he is part may not.

## 8.4   A Typology of Compliance Tactics

Based on the fundamental insights discussed above, this section will present the two-dimensional conceptual structure within which the tactics for stimulating compliance will be positioned. The horizontal dimension represents the focus of the compliance solution, for which the fundamental characteristics of both the rationalist and the normative compliance approaches are used as defining elements. As these types of theory provide different perspectives on behavioral motivations for compliance, they can accommodate tactics of a different nature. The rationalist perspective puts forward *inducement* (the use of incentives or rewards) and *enforcement* (the use of disincentives or penalties), whereas the normative perspective offers *management*[3] of compliance (the use of cooperation, assistance and persuasion). After most tactics were identified in the literature review, we also added *assessment* (the use of e.g. self-reporting, external audits and self-reflection in order to obtain insights and achieve transparency). The tactics herein were originally placed in other columns. However, since assessment is a prerequisite for both the rational and normative approaches (cf. Faure and Lefevere, 2005), and the set of management tactics grew to be relatively large, we decided to acknowledge this as a separate class. Another rationale for this is that assessments can be expected to have a deterrent effect by themselves, which constitutes another reason for them to be regarded as an autonomous group of tactics (Short and Toffel, 2010; Foorthuis et al., 2010).

---

3   We use "management" here to refer to the ways in which the normative approach encourages compliance, since this term is widely used in the literature. Note that we use "compliance management" in a broader sense, viz. also encompassing inducement, enforcement and assessment.

| Organizational Level | Inducement | Enforcement | Assessment | Management |
|---|---|---|---|---|
| Enterprise | Mandating compliance officers to provide incentives; Developing guidelines for rewarding | Mandating compliance officers to provide disincentives; Developing guidelines for punishing; Making examples out of non-compliers | Gaining insight into enterprise-wide compliance rates; Reflecting on culture in terms of compliance; Auditing the compliance function; Using self-reporting; Installing a whistle-blower hotline; Training compliance assessors; Subjecting operations to third party scrutiny*; Internal regulation by information* | Creating internal commitment by external justification; Properly specifying policies; Actively disseminating policies; Preventing non-compliance; Ensuring management support; Understanding & using social structures; Gathering and creating knowledge; Facilitating communities of practice; Socialization of newcomers; Implementing self-regulation* |
| Collective | Funding compliance-unrelated expenses; Providing political support for other initiatives; Celebrating successes | Rejecting the project deliverable; Terminating the project | Using IT-systems for storing and managing auditable information; Automatically assessing compliance; Manually assessing compliance | Providing assistance; Compensating for compliance costs; Selecting compatible and knowledgeable employees; Using internal market logic* |
| Individual | Offering formal rewards; Offering social rewards; Offering professional rewards | Imposing formal penalties; Imposing social penalties | Providing performance feedback | Training regulated actors; Employing social psychological mechanisms; Providing organizational support; Signing an agreement |

*Focus*

Figure 8.2. Typology of Compliance Tactics

The vertical dimension represents the organizational level at which the tactics are applied, i.e. the level at which the internal regulators' effort is made (note that it does not denote the actor at which the tactic is targeted). The *enterprise level* is the level at which the internal policy and its norms are formulated, perhaps based on external input. This is also the level at which the compliance management strategy is developed and at which senior management, compliance officers and organization-wide auditors operate. The *collective level* is the level of tactics for stimulating compliance of organizational units and temporary initiatives, such as projects that need to conform. The *individual level* accommodates tactics targeted at individual employees that are expected to comply.

Figure 8.2 is a visual representation of the full typology. The tactics positioned within it are discussed in detail in section 8.5 (innovative tactics are marked with an asterisk and will be discussed in 8.5.10). We will refer to the empty two-dimensional structure described above as the framework, whereas the structure including the tactics – this chapter's core contribution – will be denoted as the typology. A *typology* is used to describe conceptually derived interrelated sets of ideal types (Doty and Glick, 1994). Such types are defined by more than one attribute simultaneously (i.e. they are multidimensional by nature), with the order of the attributes being irrelevant (Marradi, 1990). A typology does not feature discrete, mathematical decision rules for classifying, and mutual exclusiveness and exhaustiveness cannot be entirely guaranteed (Doty and Glick, 1994; Rich, 1992; Smith, 2002). Our inventory of compliance tactics constitutes a typology, since they are defined by multiple attributes (the two dimensions of our framework plus any idiosyncratic characteristics to distinguish them within a given cell), are theoretically derived conceptual constructs (i.e. based on literature instead of e.g. empirical-based cluster analysis), and mathematical decision rules are not provided.

The framework can be used to position and characterize *individual tactics*. However, the framework can also be used to describe an organization's *compliance management strategy*, which is a general, integrated plan or pattern consisting of multiple tactics with the intention of achieving a satisfactory level of internal compliance. As a strategy utilizes multiple tactics, it typically covers several cells of the framework. We will return to the topic of compliance strategies in section 8.6.2. First, however, we will explicate the individual tactics of the typology.

## 8.5   Compliance Tactics

This section describes the typology's tactics, which have been identified during the literature study. Innovative tactics from the (inter)national laws and regulations level are translated to internal tactics for organizations in section 8.5.10.

The tactics are represented by ***bold italic text*** and can be found in the visualized typology presented in Figure 8.2. *Regular italic text* is (amongst others) used for showing relationships between tactics.

### 8.5.1   Enterprise Level Inducement and Enforcement

A first tactic at the enterprise level is ***mandating compliance officers to provide incentives***. A problem for those directly responsible for achieving compliance, e.g. security officers, is that they often lack the line authority over the employees in their compliance scope (Malloy, 2003; Boss et al., 2009; SIA, 2005). This means specifically that it will be very difficult for them to punish transgressors themselves. Rewarding complying employees, however, can be expected to be a less sensitive issue due to its positive character. ***Developing guidelines for rewarding and punishing*** results in enterprise-level standards, which should prevent rewards and penalties being given arbitrarily and inconsistently throughout the organization. This will increase the level of perceived fairness and consistency of the procedures (i.e. procedural justice), which is a significant determinant of compliance (Li et al., 2010; Weaver et al., 1999b; Tyler and Blader, 2005; Malloy, 2003). Making such norms and conditions explicit may also increase their perceived 'mandatoriness', which can further boost compliance levels (Boss et al., 2009). Although more difficult than the first tactic, it will still be necessary to ***mandate compliance officers to provide disincentives***, or at least introduce formal procedures to investigate possible non-compliance and contact transgressors' line managers to discuss punitive measures. In addition, ***examples can be made out of non-compliers*** by severely and visibly punishing notable transgressors. Even if the organization does not intent to e.g. fire or prosecute all employees that have committed a severe transgression, such examples may deter potential future non-compliers (cf. Paternoster and Simpson, 1996; Straub and Welke, 1998; Currie, 2008).

### 8.5.2   Enterprise Level Assessment

As an assessment tactic at the enterprise level, ***gaining insight into enterprise-wide compliance rates*** is necessary for both internal compliance management and external reporting. Important features of such a centrally available integrating analysis function are flexible reporting to different stakeholders, integration with workflows (feedback of non-compliance to those responsible and definition of escalation structures), integration with planning activities, remediation tracking and risk classification (Panitz et al., 2010; Racz et al., 2010). A central reporting function can be supported by business intelligence technology (Volonino et al., 2004; Daniel et al., 2009). Local data, such as audit trails, can be extracted and loaded in a central data warehouse, enabling aggregated reporting and holistic knowledge-development. Analytical and statistical techniques can be used to discover patterns. Standardized data models are valuable in this context, as they facilitate improved integration of local data and allow inter-organizational benchmarking (Kim et al., 2007; Panitz et al., 2010). A thorough investigation evaluates existing compliance and ethics programs, e.g. by benchmarking and using external auditors and informants (cf. Weaver et al., 1999b; Arjoon, 2006). With the insights gained, the organization can **reflect on its culture in terms of compliance**. This can be seen as a comprehensive and deep diagnosis of the corporate culture and its behavior in terms of compliance and ethics (Rossouw and Van Vuuren, 2003). This need not be an incidental affair but can be part of an ongoing process. As part of this exercise, one goal would be to gain insight into the degree of policy-induced compliance (i.e. compliance *as a result of* the compliance management approach) versus externally determined compliance (e.g. a shift in values at the societal level) (Mitchell, 1996). If compliance is the result of the chosen approach, it should be determined if actors have accepted and internalized the norms or act on rational grounds instead (Tyler and Blader, 2005). Reflecting also entails understanding non-compliance, which might be the result of high compliance costs, a poor organizational climate, a lack of technical knowledge or complex, ambiguous or difficult-to-find rules (Chayes and Chayes, 1993; Pulich and Tourigny, 2004; Zaelke et al., 2005; Mitchell, 1996). Moreover, it should be assessed whether non-compliance merely consists of transgressions or that subversion is also involved. It should also be asked how both formal internal controls and informal organizational aspects contribute to or detract from desirable compliance behavior (Weaver et al., 1999b). More fundamentally, norms and policies should be evaluated on their desirability and completeness, an endeavor that may require engagement of both internal and external stakeholders (Rossouw and Van Vuuren, 2003; Welcomer, 2002; SIA, 2005). It should furthermore be verified whether existing policies actually contain practical and valuable norms, as opposed to e.g. 'ivory tower' standards. Depending on the situation, there are also different forms, domains and methods of reflection. Risk analysis is one important form of reflection, focusing not only on risks regarding non-compliance, but also on risks in achieving the company's objectives. In the domain of security and privacy, for example, a risk analysis can use questionnaires as a method for measuring employees' attitudes towards compliance with encryption rules and identifying reasons for non-compliance (Gaunt, 1998; Hurley, 2004). Such questionnaires can also be used for more practical purposes, such as identifying lacking skills in order to specify training requirements and learn what type of new personnel to hire (ibid.). All insights mentioned above should inform the development of a new compliance management strategy. In addition, the mere act of openly discussing ethics and norms tends to encourage good behavior (Treviño et al., 2006).

*Auditing the compliance function* is a more tangible means for gaining enterprise-level insight into the current state of affairs (IIA, 1997; Hamilton, 1995; IFAC 2003; Hayes et al., 2005; SIA, 2005). Such assessments focus on entire compliance programs and on the system of internal controls that aims to guarantee compliance. Even the audit function itself can be verified, e.g. against generally accepted accounting principles. ***Using self-reporting*** of compliance status from actors may constitute an effective tactic for obtaining knowledge on compliance levels of individual actors (Heyes, 2000). Self-reporting of transgressions can directly induce self-correction, but a report can also be a motivation for having compliance verified in more detail by an independent assessor. The degree to which self-reporting is a suitable means may depend on the intentions of the actor in question (Stanton et al., 2005) and the degree of trust in the relationship (Burgemeestre et al., 2009). ***Installing a whistle-blower hotline*** ensures that employees can safely report violations internally (Paternoster and Simpson, 1996; Weaver et al., 1999a). Such a function should guarantee professional treatment of the report, and anonymity to appease fear of retaliation (Geva, 2006). Another tactic at the enterprise level is ***training compliance assessors***, such as auditors, which is necessary in order to ensure professional competence (cf. Hayes et al., 2005; IFAC, 2003). It is also desirable because it stimulates that both assessments performed and measures taken are consistent throughout the organization, increasing perceived fairness (cf. Tyler and Blader, 2005; Foorthuis et al, 2012).

### 8.5.3 Enterprise Level Management

As a crucial aspect of compliance management at the enterprise-level, ***creating internal commitment by external justification*** can be achieved in various ways. One fundamental way is to ensure the internal policy itself is consistent with the norms of the organization's environment. This not only yields external support, but also mobilizes commitment of internal participants (Meyer and Rowan, 1977). Organizations may choose, for example, to be consistent with new, prestigious or fashionable systems development approaches, project management methodologies and other industrial standards. Another way to create commitment to internal policies is to have them justified by highly professionalized external consultants. Their blessings can strengthen internal legitimacy and prevent the rise of internal dissidents (ibid.). ***Policies should also be properly specified*** in order to be effective. The mere act of specifying norms may increase their perceived 'mandatoriness' (Boss et al., 2009) and prevent 'moral laxity' (Geva, 2006). The nature of the norms should be decided for each policy.

First, the *abstraction level* should be decided upon, namely opting for high-level principles versus detailed rules (Arjoon, 2006; Burgemeestre et al., 2009). A rule-based approach assumes that actors will not hide behind the rules. A principles-based approach assumes that actors can and will reflect upon the outcomes of their actions, and that they will not abuse the freedom inherent in such an approach. A second but related decision concerns the *degree of discretion* that the regulated actor is granted, i.e. stating performance objectives versus precise methods to achieve them (Philippe and Durand, 2011; Schneiberg and Bartley, 2008). As norms are often found to be ambiguous, group-specific and context-dependent, it is crucial that they be *specified* as clearly and explicitly as possible (Mitchell, 1996; Rossouw and Van Vuuren, 2003; Foorthuis et al., 2012; Malloy, 2003). A prescription statement should be precise and detailed, albeit not too precise and detailed, since to express one thing is to exclude the other (Chayes and Chayes, 1993). A norm's rationale should also be explicitly stated. A rationale describes the reasons behind the prescription, i.e. the business benefits that can be achieved or the threats that can be averted, and should act as a motivator for compliance (Richardson et al., 1990; Emmerich et al., 1999; Pahnila et al, 2007). Explicit norms and their rationales may be especially important in cases where employees tend to have incorrect theories regarding the world (cf. Treviño et al., 2006). Norms should also be prioritized to guide decisions in case of conflicts, and examples can be provided to help with their interpretation (Geva, 2006; Foorthuis et al., 2012). Finally, policies should neither be too long nor too short (Pahnila et al, 2007). The mere existence of a policy obviously does not ensure compliance. After specifying, therefore, **the policies need to be actively disseminated** throughout the enterprise, with the intention to inform the respective population of regulated actors of the existence, content and importance of the policy, and to convince them to comply with it (cf. Zaelke et al., 2005). In order for norms to influence behavior, it is important that they are salient and 'activated' in the particular situation to which they apply (Malloy, 2003). The policy should not only be actively communicated ('push'), but also be easily accessible via e.g. the organization's intranet ('pull') (cf. Gaunt, 1998; Pahnila et al., 2007). The internal communications staff should be involved to make maximum use of existing facilities. The tone of the communication (e.g. 'tell' or 'sell') is dependent upon the audience and the nature of the relationship (Braganza and Franken, 2007). As part of the dissemination process, the internal regulators can require employees to acknowledge receipt of the policy and even to acknowledge their intention of complying with it (Weaver et al., 1999b).

  Instead of merely prohibiting certain undesired acts, they can also be **prevented** altogether, usually by restricting behavior preceding the transgressive act (Straub, 1990; Mitchell, 1990; Straub and Welke, 1998). For example, chaperoning may prevent illegal transactions, with a compliance officer accompanying and supervising crucial meetings (cf. SIA, 2005). In addition, using deontic concepts (such as permissions, obligations, prohibitions, dispensations and delegations) and technical means such as encryption and firewalls, IT-systems and data can be electronically secured from inquisitive but unauthorized actors (Cuppens-Boulahia and Cuppens, 2008; Breaux et al., 2009; Kagal et al., 2003). Another important factor in achieving compliance is ***ensuring management support*** for policies and compliance related matters. Research has shown that commitment of senior management has important implications for the scope and control orientation (e.g. rationalistic versus normative) of compliance programs (Weaver et al., 1999a). Support entails, firstly, that management actively and visibly propagates the importance of the policy and of compliance with its norms (Pahnila et al., 2007; Boss et al., 2009; Foorthuis et al., 2010; Puhakainen and Siponen, 2010; SIA, 2005). In addition to jawboning, this includes approving a policy, so as to grant it a formal status. Secondly, management should mobilize resources, by making decisions that may not benefit all local organizational units, and by providing compliance personnel with access to senior management (Boh and Yellin, 2007; Malloy, 2003; SIA, 2005). Thirdly, management should set the example in words and deeds by showing adherence to the policy, as its norms will lose credibility otherwise, leaving middle managers and regular employees discouraged to comply (Rossouw and Van Vuuren, 2003; Tyler and Blader, 2005; Puhakainen and Siponen, 2010; Pulich and Tourigny, 2004). Needless to say, a condition for providing support is that management is aware of the necessity of compliance. This could mean that they should be *made* aware by compliance officers.

As organizations are not unitary actors, it is increasingly recognized that ***understanding and using social structures*** is crucial for obtaining compliance (Zaelke et al., 2005; Malloy, 2003). For example, internal power relationships between e.g. CEO, CFO and CIO can play a role in compliance with Sarbanes-Oxley (Braganza and Franken, 2007). Promoting the policy and actual compliance are dependent on the support and commitment of key decision makers. Such support and commitment can be stimulated by gestures of goodwill. Another way of using social relationships for increasing the likelihood of compliance is empowering those key actors (Zaelke, Stilwell and Young, 2005), for example by expanding or extending their mandates or by other tactics mentioned in this chapter, such as *funding* and *assistance*. Both the selection of such tactics and the specific implementation of a given tactic may depend on the nature of these existing social relationships. Good informal personal relationships should also be mentioned here, as they allow for better cooperation, discussions on compliance and fine-grained insights into the actors' interests and reasons for non-compliance (cf. Wasserman, 2009).

*Gathering and creating knowledge* can be valuable for several reasons (Braganza and Franken, 2007; Straub and Welke, 1998). First, as the semantic, conceptual, technical, procedural and judicial aspects of external policies can be quite complex, the organization needs to prepare itself for policy implementation. This entails collecting and understanding not only those policies and their norms, but also any additional sources that can help with their interpretation. Secondly, the organization can explore academic knowledge and orient itself to generic practitioner reference models for compliance. Examples of the latter are Red Book 2.0, i.e. the GRC Capability Model (OCEG, 2009) and frameworks for auditing (e.g. IFAC, 2003). There are also frameworks for general governance (COSO, 1994) and IT governance, such as COBIT (IT Governance Institute, 2000) and SAC (Mair, 2002). Entire reference models may be implemented or only specific elements may be used. A combination of approaches may also be employed, as their focus is often complementary (Luthy and Forcht, 2006). COSO has a broader scope, for example, but COBIT provides more detailed specifications. Thirdly, the enterprise needs to build knowledge on how best to apply and deal with the norms, given its own unique characteristics and context. It is likely that this knowledge has to be created from practical experiences within the organization itself. Fourthly, external policies may not only constrain a business organization, but may also provide competitive opportunities and drivers for innovation. Compliance often entails keeping business records and process logs (see also the *gaining insight into enterprise-wide compliance rates* and *using IT-systems for storing and managing auditable information* tactics). When dealt with from a holistic framework, this can yield improved insights into the entire organization. Such insights can be used not only for taking corrective action, but also for improving overall efficiency and the organization's competitive position (Volonino et al., 2004). More specifically, they provide opportunities for redesigning processes to reduce complexity and redundancy, and to improve information quality and risk management (ibid.; Braganza and Franken, 2007; Racz et al., 2010). Once a compliance knowledge base is built, its contents should be made available. A central repository, such as the company's intranet (Kerrigan et al., 2003; Braganza and Franken, 2007), provides a necessary but passive way of knowledge dissemination. Therefore, it is likely that more active complementary activities are required, such as *training*.

Another significant way of disseminating knowledge is by **facilitating communities of practice**, in which practitioners can discuss and share explicit, tacit and cultural knowledge and norms (Wenger, 2000; Choo, 2006). Such communities can organize knowledge exchanges, by which the different types of experiences and knowledge mentioned in *the gathering and creating knowledge* tactic can be communicated, negotiated and made more explicit.

Communities of practice, therefore, enable both creation and dissemination of knowledge. Moreover, being social communities, they can help promote, institutionalize and internalize norms, create commitment, and as such encourage compliance (cf. Daniel et al., 2003). Such communities are largely self-organized and informal in nature. However, although management need not formally control them, communities may be facilitated by providing them with resources, time and freedom.

A related tactic is the institutionalized *socialization of newcomers*, which is an enterprise-wide means to facilitate newcomers' learning about desirable and effective organizational behavior and to promote a culture of compliance. It is associated with the development of positive attitudes, such as higher job involvement, organizational commitment and compliance with organizational standards (Pulich and Tourigny, 2004). In this regard, newcomer orientation provides a way to communicate organizational norms and values (and why they are deemed important), the organization's mission and vision, and its appraisal and reward system. This can be communicated through *training*, mentorship programs, role modeling, corporate videos, welcome sessions and induction programs (ibid.; Treviño et al., 2006; Hallier and James, 1999). This should increase newcomers' sense of belonging, integrate them within the organizational culture, and teach them the values, norms, beliefs and preferred behaviors. Instead of leaving it entirely to HR, managers should be actively involved in newcomer orientation (Pulich and Tourigny, 2004) and promote a culture of compliance (SIA, 2005). (See the *ensuring management support* tactic for other ways in which management can contribute.) Socialization processes may take the form of organizational rituals, ceremonies, storytelling and other symbolic acts (Islam and Zyphur, 2009; Thralls, 1992; Pulich and Tourigny, 2004). However, rituals and ceremonials can also have a counterproductive effect, such as role distance and instrumental attachment (Hallier and James, 1999; Islam and Zyphur, 2009). Examples of rituals are *rites of passage* (consisting of symbolic acts to allow employees to transition to a new role and normative setting) and *rites of renewal* (consisting of symbolic acts that are periodically staged to reassert the dominance of certain organizational values) (Islam and Zyphur, 2009; Thralls, 1992).

### 8.5.4 Collective Level Inducement and Enforcement

As an incentive at the collective level, certain *compliance-unrelated expenses might be funded*, as a reward for compliance with the policy. The IT costs of a project, for example, are sometimes centrally funded on the condition that it conforms to the enterprise architecture (Foorthuis et al., 2010, 2012). In addition, compliance can be rewarded by *providing political support* for other initiatives of the actor in question, such as an approval of ambitious local programs or support for the appointment of a department's senior representative at an influential position within the organization. Another tactic is the organization-wide *celebration of successes* if they are achieved in a compliant fashion. Publicly acknowledging successes not only fosters team coherence, but also values that are intrinsic to high performance (Hefner and Malcolm, 2002). This will demonstrate to other actors that compliance need not be a restriction and may even increase chances of success.

A project or department can be punished for non-compliance by *rejecting the project deliverable* (Project Management Institute, 2004). This can occur, for example, if a developed software solution has not been described in sufficient detail to meet the standards set by the party responsible for maintaining it in production (especially if this is an external party). Rejection need not be final, since the deliverable may be accepted after reworking it in accordance with the norms. However, in extreme cases, *terminating the project* may be a serious alternative.

### 8.5.5 Collective Level Assessment

*Using IT-systems for storing and managing auditable information* ensures verifiability (Currie, 2008; Daniel et al., 2009). Compliance often entails keeping business records, so as to facilitate standard control and reporting processes and to ensure the possibility of future assessments. Although systems at the enterprise level will in practice also store records (e.g ERP-systems), keeping log trails mostly involves business applications of specific organizational units. In general, systems need to be able to store time-dependent data, offer version-control, register employees updating information, authenticate data (e.g. checksum functionality), enable fine-grained permanent deletion of privacy-sensitive data, and back-up and archive data with third parties (Peterson and Burns, 2006). Logging process and decision trails in IT-systems renders actors traceable, the awareness of which may have a direct deterrent effect and makes this a tactic in its own right. However, trails storage is also a prerequisite for performing assessments. Compliance assessments of log trails and other artifacts of decisions, processes, systems and projects are conducted to verify whether the norms are actually complied with in practice (Ellis et al., 1993; Botha and Boon, 2003; Foorthuis et al., 2012; Agrawal et al., 2006). A recent survey ($n$=293) found the use of compliance assessments to be the most important determinant of conformance to organizational standards, probably due to assessments emphasizing the importance of compliance and to actors' desire to avoid confrontation (Foorthuis et al., 2010). The results of such an assessment or audit can be reason to take corrective action. The object of scrutiny here can be behavior, such as when it is verified whether a project conforms to the standards of the relevant project management or systems development methodology. In addition, an assessment can verify product quality by reviewing the project's design documents or by checking the delivered output of a production process against the quality standards. As a result of our definition of compliance, what is central in an assessment is whether the behavior and products are consistent with the norms, not if they are consistent *because of* the norms. Finding the actual causes of compliance is explored in the aforementioned *reflecting* tactic. Several types of assessments can be acknowledged, which will be discussed below.

*Automatic compliance assessments* determine electronically whether an organization and its processes and data conform to the norms. Such a verification process uses formal process models, meta models, rules and constraints. A distinction can be made between design-time (e.g. when designing business processes), run-time (when actually running business processes) and after-the-fact (when verifying audit trails) compliance checking (Ly et al., 2008). Research has been conducted on implementation in various technologies, such as Formal Contract Language, First-Order Predicate Logic, Concurrent Transaction Logic, Event Algebra, Concurrent Temporal Logic and Linear Temporal Logic (Governatori et al., 2006; Kerrigan et al., 2003; Ly et al., 2008; Kim et al., 2007). Automated verification offers large coverage – due to fast automated checking – and accurate and consistent results – due to the required formalization of norms (El Kharbili et al., 2008; Hurley, 2004). This type of surveillance is therefore well suited for continuous routine monitoring. Automated systems may also assist managers in organizing written assessment reports, help assign responsibilities to employees, and enforce procedures instead of merely verifying them (Agrawal et al., 2006). Employing automated solutions therefore has the additional benefit of freeing up resources that can be reallocated to other purposes (Hurley, 2004). Despite these benefits, however, *manual compliance assessments* will remain necessary. Not all norms are suitable for automated checking, not only as a result of inherently abstract and multi-interpretable prescriptions stated in natural language, but also because of the importance of sense-making and personal and contextual knowledge in the verification process (Chayes and Chayes, 1993; Foorthuis et al., 2012). Automatic compliance assessments therefore may often result mainly in self-evident outcomes. Conducted by human agents, manual assessments can on the other hand be used for sophisticated, subtle, subjective and situated evaluation. This may be especially relevant in specific cases in which an in-depth audit is required. Examples of methods that can be employed are interviewing employees, observing operations, and reviewing records and documents (ibid., Short and Toffel, 2010).

### 8.5.6 Collective Level Management

***Providing assistance*** to projects or organizational units can increase compliance levels. For example, technical and personnel assistance can be provided to ensure that norms are explained and that sufficient guidance, knowledge and capacity are available (Braganza and Franken, 2007; Zaelke et al., 2005; Foorthuis et al., 2010). Individuals can also be offered help and assistance when confronted with ethical and compliance issues, e.g. by a telephone advice service (Weaver et al., 1999b). Such facilitating conditions may also positively impact attitudes towards complying (Pahnila et al., 2007). Knowledge can relate to the norms directly (helping with their interpretation) or to required technical expertise. As a specific form of assistance, ***compliance costs can be compensated for***, so as to let compliance be a budget-neutral aspect of business (Mitchell, 1996; Braganza and Franken, 2007). Apart from accessible *knowledge repositories*, *facilitating communities of practice* and *training*, knowledge can be brought in by ***selecting compatible and knowledgeable employees***. Selecting here refers not only to hiring personnel from outside the organization, but also to matching internal employees with projects, and moving individuals between subsidiaries on secondment (Harris and Cummings, 2007; Braganza and Franken, 2007). Critical selection is also important when hiring or appointing employees that have to make ethics and compliance-related decisions in ambiguous situations (Treviño et al., 2006). Selection criteria are not restricted to specific knowledge (resulting from relevant education or working experience), but also involve congruent and accepted ethical standards (ibid.; Tyler and Blader, 2005; Geva, 2006; DiMaggio and Powell, 1983) and the expectation of whether candidates will abide by the organizational rules (Pulich and Tourigny, 2004; Ogbonna and Harris, 1998). In the case of external consultants a welcome bonus may be that they are less biased and are unhampered by the organization chart (Burditt, 1996). During the selection process, compliance personnel may go as far as to check a candidate's disciplinary or complaint history (SIA, 2005).

### 8.5.7 Individual Level Inducement and Enforcement

At the individual level, ***offering formal rewards*** is a well-known inducement tactic. This can take a variety of forms, such as pay raises, promotions, awards, bonuses, days off, paid vacations, sabbaticals and stocks (Bulgurcu et al., 2010; Stajkovic and Luthans, 1997; Paternoster and Simpson, 1996). Interestingly enough, an organization should not blindly offer incentives. In the case of employees who have already internalized the respective norms, rewards may work counterproductive, as providing incentives to act morally might undermine the intrinsic value of ethical behavior or might lower the responsibility felt by individuals (Treviño et al., 2006; Brekke et al., 2003). ***Offering social rewards*** focuses on proven informal incentives, such as compliments, praise and commendations, a word of thanks, personal mention and appreciation in oral or written assessments, celebrating heroes, and the increase in one's self-respect, reputation and status that come with that (Stajkovic and Luthans, 1997; Rossouw and Van Vuuren, 2003; Bulgurcu et al., 2010; Pulich and Tourigny, 2004). In addition, adherents can also be ***offered professional rewards***, such as interesting courses or challenging and prestigious assignments.

There are also several tactics related to punishment at this level. Ample research has been carried out on penalties, often from the perspective of deterrence theory. Severity, certainty and celerity have been shown to be crucial factors in this context (Paternoster and Simpson, 1996; Tyler and Blader, 2005; Bulgurcu et al., 2010). ***Imposing formal penalties*** brings with it threats such as monetary or legal sanctions, demotion, loss of occupational position and jeopardizing future job prospects (Paternoster and Simpson, 1996; Bulgurcu et al., 2010; Straub and Welke, 1998). Usually, an organization will make use of 'progressive discipline', in which oral warnings are followed by written warning, suspension and finally job termination (Pulich and Tourigny, 2004). Another type of disincentives for individuals is ***imposing social penalties*** on violators. These more intangible tactics might take the form of reprimands, 'naming and shaming', suspension, unfavorable mention in oral or written assessments and the consequent loss of self-respect, reputation and status (Malloy, 2003; Braganza and Franken, 2007; Bulgurcu et al., 2010; Paternoster and Simpson, 1996).

### 8.5.8   Individual Level Assessment

***Providing performance feedback*** is a proven and inexpensive management tactic for improving employee behavior (Kluger and DeNisi, 1996; Stajkovic and Luthans, 1997; Pulich and Tourigny, 2004). This tactic derives its power from providing the employee with objective information on his performance, preferably presented in an immediate, positive and specific fashion. The focus should be on the task rather than the person. Examples of desired behavior can also be provided. The information can trigger a performance-improving reaction within the employee, for example because he is encouraged to reduce the discrepancy between his performance and the standard or because of an inner motivation to raise the bar.

### 8.5.9   Individual Level Management

***Training regulated actors*** clarifies the rules and the measures taken against transgressors (Mitchell, 1996). In the context of security policies, for example, a training should communicate and spread understanding of the norms ("encrypt e-mails"), provide required knowledge (information classification rules) and teach necessary skills (using encryption software) (Puhakainen and Siponen, 2010; Braganza and Franken, 2007; Straub and Welke, 1998). More in general, an important goal is making employees aware of the prescriptions' underlying rationales and benefits, and of the consequences and risks of non-compliance (e.g. making the organization vulnerable to an outside attack). This awareness should influence the attitudes of employees and the consequent intentions to comply. Communicating the penalties and rewards during the training may likewise influence attitudes (Bulgurcu et al., 2010; Straub and Welke, 1998). Organizations should organize frequent compliance trainings (MacLean and Behnam, 2010). Furthermore, different types of training may be required for different types of employees. For example, technical people may need less training on encryption tools than do sales people (Puhakainen and Siponen, 2010). In order to increase motivation, training should also focus on tasks relevant to the course participants (ibid.).

*Employing social psychological mechanisms*, in essence, consists of a wide variety of compliance techniques to influence individuals (Zimbardo and Leippe, 1991; Cialdini and Goldstein, 2004). For example, the *door-in-the-face* technique involves making a large request, which is likely to get rejected. The actor may subsequently reciprocate the requester's 'concession' by his or her own concession, i.e. complying. When the *that's-not-all* technique is employed, the target is presented with an initial request, after which a bonus is added to the deal. The increased benefits, perceptual contrast and the urge to reciprocate often result in compliance. The *foot-in-the-door* technique works by first asking the compliance-target to comply with a small and easy request. After compliance is assured, a larger and often related request is made, resulting in compliant behavior. Another technique here is to use the *obedience to authority* rule, which seems deeply ingrained in human nature. This rule makes it very hard to say "no" to authority figures, even if the request would inflict damage to others. A final technique we will mention here is *good cop/bad cop*, in which one professional takes a harsh approach in trying to have the actor comply. His colleague subsequently takes a soft and conciliatory approach, pretending to be allied with the actor and promising to prevent the severe penalties the bad cop threatened him with (Cialdini, 1987). See the social psychological literature for more techniques. The organization could also have personnel **sign an agreement** (Gaunt, 1998; Straub, 1990). Working with confidentiality agreements, for example, is common when employees have to work with sensitive data. Explicitly agreeing with a policy may increase the commitment felt by the regulatees, as they are given responsibility and accountability. Note that all employees have a 'psychological contract' with their organization, taking the form of expectations regarding the duties of the organization and of themselves. Such expectations tend to be implicit, so managers should try and clarify them (Pulich and Tourigny, 2004). Another managerial tactic is ***providing organizational support***, in which employees receive professional help when dealing with personal problems that obstruct good functioning (Pulich and Tourigny, 2004). Some deviant behaviors may be dealt with directly, e.g. by anger management programs or other types of correctional counseling. However, support may have a more indirect character, for instance when the employee is confronted with problems such as divorce, sickness or loss of a family member. Providing support may not only help the individual employee, but can also be of value to the organization by preventing escalation at work.

### 8.5.10 Tactics and Innovations in Compliance

Driven by regulatory crises, new visions and rulemaking experiments, scholars in law, sociology and political science have defined innovative forms of regulation at the (inter)national level (Schneiberg and Bartley, 2008; cf. Faure and Lefevere, 2005; cf. Zaelke et al., 2005). In the organizational sciences, however, daring and innovative new tactics seem to be few and far between (an obvious exception being the use of enabling IT to assess and control compliance, as described in the *automatically assessing compliance* and *preventing non-compliance* tactics).

At the (inter)national level, new and innovative forms of regulation may have one or several of the following characteristics (ibid.). First, although traditional schemes often relied on state agencies for regulation and enforcement, new forms often also rely on non-state actors such as NGOs to act as watchdogs and assert pressure. Secondly, although regulation was traditionally used to suppress competition, some new forms have embraced it as a means to achieve regulatory goals. For example, tradable permit systems set a maximum level of allowed pollution, subsequently issue permits amounting to that level, and let firms choose their individual pollution levels by having them trade those permits. Thirdly, instead of detailed and universal prescriptions, new regulatory forms increasingly stimulate self-regulation of 'regulated' actors. For instance, those actors can define their own standards through collective dialogue. Such processes may be driven by state-defined high-level goals, but allow for individual freedom. Fourthly, whereas traditional forms made extensive use of formal sanctions in an enforcement approach, new forms often rely on information-based schemes in which rating and ranking, naming and shaming, peer reviews and benchmarking, price-driven processes, and empowered non-state actors play a crucial role. Sanctions are not an integral part of the regulation, nor are they administered by the state. Given actual transparency, rewards and penalties in such an approach are instead invoked decentrally by consumers, investors and advocacy groups in terms of products consumed, investments made and reputations defined. A California law, for example, resulted in improved behavior by forcing restaurants to display health inspection letter grades in their windows (Schneiberg and Bartley, 2008).

Although we could not find them in the organizational sciences literature, the analogues (or translations) below can be defined as candidate tactics for internal compliance. They are marked with an asterisk in Figure 8.2.

- *Subjecting operations to third party scrutiny*. If an organization sincerely strives to be in compliance, it may motivate its members by calling upon external parties. For instance, a large energy firm could provide an NGO with access to its plants in order to have operations scrutinized. This would not only ensure critical compliance assessments, but society's knowledge of such an endeavor would also have positive effects on the company's reputation. The involved parties should make strict agreements regarding public disclosure of the results, in order to allow the company sufficient time to address any violations. Another expected condition for this approach, in which external parties are granted access willingly, is that the firm be reasonably confident about receiving positive evaluations.
- *Using internal market logic*. Competition can be introduced within an organization for obtaining compliance-related goals. For example, by letting multiple teams draft a proposal for a business project that has to implement or comply with e.g. regulations or internal IT standards. Management can subsequently decide which proposal offers the optimal solution, taking into account both business goals and policy constraints. Note that we positioned this tactic at the Collective level of the typology as a result of our example implementation. However, a spin-off version of this tactic could possibly be applicable at the Enterprise level.
- *Implementing self-regulation*. Self-regulation could be implemented within the organization by an enterprise-wide policy containing high-level targets and abstract principles, allowing actors to conform to them by translating these principles to their own explicit and more detailed rules, according to local circumstances. It would be crucial that enterprise-wide regulators restrict themselves to guarding that local units conform to the high-level goals of the policy (e.g. selling mortgages ethically) and achieve its targets (e.g. having a tiny percentage of clients with difficulties paying off their debts). This is a consequence of the self-regulation concept, using abstract principles and focusing on end goals. The internal regulators therefore cannot readily assess compliance with their principles, since these are too abstract and the norms that are sufficiently tangible for assessments are owned by the 'regulatees' (e.g. local offices). Opportunism and free riding are evident risks in a self-regulatory scheme. However, such risks may be relatively manageable within an organization by asserting internal pressure. The organization's regulators would also be able to review and approve the self-written tangible local norms before implementation begins. Moreover, they would assess the achievement of the end goals. A wholly different approach to internal self-regulation would be to still create detailed rules at the enterprise level, but regularly rewrite them in cooperation with regulated actors as knowledge of and experience with applying those rules is build up.
- *Internal regulation by information*. Regulation by information could be implemented by internally publishing information about actors' compliance performance. For example, if compliance assessments of projects or units with internal IT standards are conducted, the results could be presented on the organization's intranet or in a newsletter. As actors' reputations are at stake, such disclosure may have a significant effect on compliance results, especially when the most important norms are widely accepted. Compliance reporting of individuals may be more difficult. However, conforming behavior may be communicated in a positive manner, including any bonuses these individuals may have received as a reward for it. This approach may not be a single tactic, but rather a mix of tactics drawn mainly from the Assessment column and may also be combined with self-regulation. Another option would be to completely dispense with creating explicit norms, as transparency could lead to the organizational community determining acceptable levels of behavior by itself.

## 8.6  Discussion

In this section we will discuss the results of our literature review. Subsection 8.6.1 presents the general findings. Subsection 8.6.2 focuses on developing coherent compliance strategies, which is relevant because a fragmented set of tactics is not sufficient for effective compliance management.

### 8.6.1  General findings

We start our discussion with a short overview of each academic field's contributions to this study. The fields of *law* and *political science* prove to have the most mature conceptualizations of compliance and related concepts. These insights therefore figure largely in section 8.3. The reviewed literature in the field of *philosophy* was also used mainly for the foundational insights presented in that section. The organizational sciences – such as *management, business studies, information systems* and *organizational sociology* – were used mainly for the identification of tactics and insights into strategies. As stated in section 8.5.10, except for IT-related solutions, the organizational sciences do not seem to report on many innovations and experiments in compliance. It is mainly the fields of law, sociology and political science that have described innovative forms of regulatory implementation at the (inter)national level. Translating them to the organizational domain, we have defined four analogues of innovative candidate tactics for internal compliance.

Interestingly, during our literature study we practically did not find any publications presenting a general and comprehensive perspective on organizational compliance strategies. Note that the compliance strategy item in Table 8.3 does not refer to publications that put forward such rich conceptualizations of compliance strategies, but to publications that in some way present input for such a view (as such, they will be referenced in section 8.6.2 on strategies). Most publications presenting organizational compliance solutions do not focus on general insights into compliance strategies, but instead aim to make the case for a specific compliance view or approach, employing a selected set of tactics (e.g. Tyler and Blader, 2005; Weaver at al., 1999a; Panitz et al., 2010; Puhakainen and Siponen, 2010; Agrawal et al., 2006; Straub, 1990; Li et al., 2010; Boss et al. 2009; Namiri and Stojanovic, 2007; Foorthuis et al., 2010; Governatori et al., 2006; Breaux et al., 2009; Peterson and Burns, 2006; Cuppens-Boulahia and Cuppens, 2008; Straub and Welke, 1998; Racz et al., 2010; Bulgurcu et al., 2010; Kerrigan et al., 2003; Kim et al., 2007; Pahnila et al., 2007). Although some publications do present a neutral, non-evaluative and broad overview of compliance tactics available, the number of identified tactics tends to be quite limited (Braganza and Franken, 2007; Pulich and Tourigny, 2004; Weaver et al., 1999b; SIA, 2005). Moreover, only a limited number of publications have provided an integrated or high-level discussion on the concept of compliance strategies and programs (e.g. COSO, 1994; Luthy and Forcht, 2006; Burgemeestre et al., 2009). Notable in this regard is Weaver's et al. (1999a) view of ethics programs as organizational control systems. Such programs are characterized by a control orientation (enforcement versus management), have a scope in terms of formalization, specialization and hierarchy (in terms of e.g. ethics officers and staff), and consist of other elements such as the ethical codes of conduct, communication systems, training programs and disciplinary measures. However, Weaver et al. still present a limited number of tactics and focus only on formal initiatives. Furthermore, the bird's-eye view of our literature study yielded insights that are quite distinct from each other and sometimes even incompatible. A still broader conceptualization, allowing for all kinds of approaches, is therefore required. Section 8.6.2 aims to present such a rich conceptualization of compliance strategies.

Another finding concerns cell volume. Figure 8.2 makes clear that the Assessment and especially Management columns are abundantly filled, whereas the Inducement and Enforcement columns do not contain as many tactics. This could be the result of the arbitrary nature of typologies and the subjectivity of the research. However, studying the assessment and management tactics in more detail makes clear that this is not the case. Many of these individual tactics – such as *socialization of newcomers, selecting employees, facilitating communities of practice* and *reflecting on culture in terms of compliance* – offer rich opportunities and choices when implementing them (see section 8.5 for details). In other words, the assessment and management tactics are not defined with finer granularity than their inducement and enforcement counterparts. Rather, it should be concluded that the assessment and management approaches offer a wider variety of ways for encouraging compliance. Another aspect concerns the vertical dimension, with the Enterprise level accommodating relatively many tactics. Note that the tactics at this level are not necessarily employed directly by senior managers, since *gaining insight into enterprise-wide compliance rates* and *properly specifying policies*, for example, will be executed by staff members. However, senior management should be aware of their responsibility to initiate tactics at this level.

### 8.6.2 Moving From Compliance Tactics to Compliance Strategy

In section 8.5 we have described a large number of compliance measures. Although this typology of tactics constitutes the core contribution of this chapter, a fragmented set of tactics is not sufficient for effective compliance management. Therefore, this section aims to present a rich conceptualization of compliance management strategies and discuss crucial aspects of developing such a coherent strategy. Not surprisingly, in most organizations it will first be necessary *to put the compliance management function in place*. This may encompass the installment of an independent audit and compliance department – including its officers and staff with clearly defined roles, responsibilities and authority (Gaunt, 1998; Weaver at al., 1999b; Arjoon, 2006; SIA, 2005). Depending on the situation, the scope of such initiatives tends to vary widely in terms of personnel and supporting structures (Weaver at al., 1999a; SIA, 2005). An important implementation aspect is the necessary separation of duties, so as to guarantee independent and objective assessments and audits (Solms, 2005; SIA, 2005). In addition, employees occupying relevant positions should be provided with proper mandates, authorizations and authority (Agrawal et al., 2006; Arjoon, 2006; COSO, 1994).

A key responsibility of the compliance function is the development of a compliance management strategy. As stated, however, we did not find any publications presenting a general and all-encompassing perspective on organizational compliance strategies. The remainder of this section aims to present such a conceptualization. We will draw on the general strategy research of Mintzberg (1987) for this purpose. There are several reasons for using Mintzberg's five views on strategy. First, the concept of organizational strategy is ambiguous and multifaceted (French, 2009), and Mintzberg's comprehensive view also allow for perspectives on strategy that are incompatible for a given situation (e.g. a predefined plan versus a pattern discovered after the fact). Second, Mintzberg's conceptualization is widely accepted.

The most dominant view is that of strategy as a ***plan***, a consciously intended course of action. The compliance strategy here is therefore developed purposefully and in advance of execution. Goal-setting will be an important element (Quinn, 1996), especially regarding the level and nature of compliance with policies in question. Note that striving for a 100% compliance level may not only be unrealistic, but could also yield unwanted side-effects. The organization would run an increased risk of developing a bureaucratic culture of 'overconformity'. Adherence to the rules would then become an end in itself rather than aim at achieving goals, resulting in rigidity and an inability to adjust to individual cases (Merton, 1957). Employees will consequently hide behind the rules and absolve themselves from any responsibility. In addition, a strong focus on compliance and monitoring may lead to a 'surveillance culture', distrust, a distinction between 'them' (i.e. the compliance officers) and 'us', a lack of commitment and even resistance (Tyler and Blader, 2005; Currie, 2008; Short and Toffel, 2010). As has been suggested in political science, an acceptable level of overall compliance may therefore be better than a standard of strict compliance (Chayes and Chayes, 1993). The organization may only be able to choose such levels in the case of standards and internal procedures, whereas in the case of laws and regulations full compliance is required. Even regarding regulatory compliance choices have to be made, however, as it may be possible and desirable to move "beyond compliance" and do more than strictly necessary (Short and Toffel, 2010). Organizations have on the other hand also been known to merely act as symbolic compliers, instead of taking substantive action (see below). In addition to determining the degree of compliance, it should also be decided for each policy *how* to comply. Organizations often have some leeway as to the exact nature of achieving compliance goals, even with external norms (Philippe and Durand, 2011). For example, organizations can comply with transparency laws (e.g. disclosing their investments) but refuse to conform to underlying norms and values of ethical business (and e.g. continue their investments in the weapons industry). In addition to goal-setting, more mundane decisions have to be made regarding *budget*. Pursuing a rationalistic command-and-control approach, with a heavy focus on monitoring and enforcement, is a very expensive road to travel (Tyler and Blader, 2005). In addition to budget, *time* is also an important aspect. Although strategy is not associated with short-term thinking, setting out to change the organization's compliance culture, for example, may require an extensive and long-term change initiative with repeating reinforcement of norms and an uncertain outcome (Gaunt, 1998; Ogbonna and Harris, 1998). The combination of compliance goals and available time and budget may require *goal prioritizing*. Schraaf (2005) describes a 2x2 matrix for setting priorities on the basis of the dimensions risk (in terms of negative impact) and non-compliance rate. Issues with both a high risk (e.g. high fines) and high internal non-compliance rate receive the highest priority.

The second view of strategy is that as ***perspective***, a way of perceiving the world. This is strongly related to the organization's worldview, culture and 'character'. An important aspect here is that strategy is a shared perspective, part of the organization's collective mind. Especially relevant in our context is what view the organization has regarding the nature of compliance. What account does it have of why its units, projects and employees comply or do not comply with policies? Are organizational actors rational agents calculating the benefits and costs of compliance decisions or are they good-faith actors valuing their identity and in need of effective means and knowledge? Or is the worldview a combination of both? This perspective, whether conscious or not, will guide strategic and day-to-day compliance decisions.

A third view is strategy as a ***ploy***, a maneuver to outwit a competitor or opponent. In terms of compliance, the organization may try to attract investors by being the first to adhere to the new rules of Sarbanes-Oxley or the Basel accords, or to innovative industrial standards. In contrast, the organization may also choose to lower compliance levels temporarily in order to achieve a strategic and highly visible business result. A ploy in the context of compliance could also be taken to mean obfuscating non-compliance (Heyes, 2000) and conducting mainly symbolic compliance activities (Short and Stoffel, 2010; Weaver et al., 1999b; Meyer and Rowan, 1977). Such symbolic compliance refers to the concept of *decoupling policy from practice*, a form of window-dressing in which an insignificant compliance program is implemented. Such programs are ineffective because they are disconnected, i.e. the organization avoids their crucial elements – such as monitoring and punishing – being integrated into the central, task-related business activities. This results in non-compliance, as processes are not continuously monitored for compliance, processes are not seriously affected by the program, and transgressors are not confronted with disciplinary measures. However, to the outside world public claims are made about significant compliance efforts and the organization being in compliance. Research has shown that organizations may get away with decoupling, but that this can also lead to severe financial and reputational damage (MacLean and Behnam, 2010). As stated earlier, in this chapter we will assume an organization genuinely strives to be compliant.

A fourth view sees strategy as a ***position***, a means of locating an organization in its environment and a mediating force between the organization and its external context. The compliance strategy in this view is used to position the enterprise in the market, for customers and other stakeholders, and in terms of which ethical principles to adhere to (and to what degree). It should also be decided which professional standards and quality norms to comply with, and which to demand from e.g. suppliers. This also points to the fact that the organization has an active relationship with its environment and should determine whether or not to try and influence external regulators' policy making processes (cf. Meyer and Rowan, 1977; Schneiberg and Bartley, 2008). Indeed, one painless way to be compliant is to ensure newly devised norms are consistent with existing behavior (Downs et al., 1996). Apart from this external orientation, the compliance strategy should also be positioned internally. It should, in other words, be aligned with the overall business strategy and with the core beliefs and values of the organizational culture. In this regard, senior management should make clear what the *purpose* of the organization is. Purpose could be regarded as the statement of an organization's moral response to its broader defined responsibilities, not as an amoral plan for exploiting commercial opportunities (Bartlett and Ghoshal, 1994). Several questions should also be considered regarding the ambitions of integration. One question concerns whether holistic compliance is desirable (Volonino et al., 2004; Cleven and Winter, 2009; Panitz et al., 2010), i.e. the extent to which compliance management intends to deliver central and integrated (instead of fragmented) compliance efforts, covers multiple internal and external policies, and has a long-term scope. Another issue is that it should be decided whether compliance management itself should be integrated with enterprise risk management and overall governance efforts (Dawson, 2008). As described in a previous section, integrated, holistic compliance may yield improved insights, which can be used for increasing consistency and reducing complexity of processes and systems.

When strategy is seen as a ***pattern***, the focus lies on the pattern that can be observed in a stream of actions. This fifth view focuses on two important aspects of compliance strategies. The first is that strategy as a plan is neither sufficient nor necessary, as there should be *actual behavior* in the form of decisions and actions. In our context this implies that day-to-day business processes are not decoupled and shielded from, but actually integrated with and affected by compliance-related considerations (MacLean and Behnam, 2010). The use of informal tactics may also be more immanent in these day-to-day behaviors than in predefined plans. A second important aspect is that of *coherency* in those decisions and actions (Mintzberg, 1987; Quinn, 1996). A compliance strategy should focus on the interdependence and internal consistency of goals, planned action and actual conduct. Moreover, and arguably the most crucial part in the context of this study, a compliance strategy should encompass one or more coherent sets of tactics aimed at achieving a targeted compliance level. The framework and typology presented in section 8.4 can be used to predefine (or analyze in retrospect) such sets. Given the fact that the typology is the core contribution of this chapter, it is this aspect that is especially interesting here. The remainder of this section will therefore focus on the factors involved in creating these intervention mixes, i.e. relevant and coherent combinations of compliance tactics.

In general, a strategy drawing its tactics from both the rationalist and normative perspective seems to be the most promising approach (Grossman and Zaelke, 2005; Wasserman, 2009; Tyler and Blader, 2005; Short and Toffel, 2010; Paternoster and Simpson, 1996; Braithwaite, 2007). From the rationalist model, especially enforcement is often seen as a necessary element. According to Grossman and Zaelke (2005, p. 78), for example, the "proper balance of the two models thus seems to be a compliance enforcement system that also encourages the norms and incentives that lead to voluntary compliance, while maintaining the bedrock foundation of enforcement and deterrence to alter the calculations of those less inclined to voluntarily comply." In terms of the framework of section 8.4 this means minimally picking tactics from both the Enforcement and Management columns.

However, a compliance strategy is heavily contingent upon the situation. Most fundamentally, the choice of which exact combination of tactics to include depends strongly on whether a rationalistic or normative orientation weighs more heavily. In this context, several strategies or "modes" can be identified (Rossouw and Van Vuuren, 2003). For example, a primarily rational "compliance mode" would focus on formalizing norms, monitoring compliance-performance and sanctioning of transgressions. Tactics would be drawn mostly from the Enforcement and Assessment columns. A primarily normative "integrity mode" would focus on reflecting on the culture, top management support, internalization of values, providing assistance and other forms of managing non-compliance. Tactics would thus be chosen mostly from the Management and Assessment columns. Note that the choice of such a mode or strategy depends not only on ambitions, but even more so on the starting point, i.e. the current compliance culture, routines and context of the organization (ibid.; Short and Toffel, 2010). As stated in the previous section, it should also be noted in this context that the Management and Assessment columns offer a wider variety of tactics, and thus more opportunities for tailoring the strategy to the specific setting.

However, it is too simplistic to presume that a single combination of tactics in the strategy is sufficient. It may very well be necessary to define intervention mixes at a lower level. For example, each *reason for non-compliance* may have its own set of tactics, as does each type of actor (Schraaf, 2005). In addition, each *geographical establishment* may have its own unique intervention mix, since organizational cultures and legal requirements differ between locations (cf. Arjoon, 2006). Furthermore, different *policies* may require different sets of compliance tactics, as compliance with internal standards presumably allows more discretion than compliance with mandatory external legislation. Internal regulators should also think about how to *sequence* or escalate the use of tactics for distinct types of situations. An approach known as 'responsive regulation' (Ayres and Braithwaite, 1992; Braithwaite, 2007) states that in most cases regulatees do not have to be coerced into complying and can be persuaded through e.g. explanation and education. However, for actors demonstrating increasing (risk of) non-compliance, an escalation path can be followed with increasingly intrusive and coercive measures taken. Such an approach, therefore, differentiates between different types of actors, risks or behavior, and deals with each accordingly. It is exactly the perception that a regulator can and if necessary will resort to varying degrees of tougher enforcement tactics that allows him to be 'soft' to most actors most of the time. Such a strategy could then draw its tactics from all columns of the framework. For some actors management tactics would suffice, whereas for others a combination of enforcement and assessment tactics would be necessary. It should be noted, however, that a compliance strategy should not try to include as many tactics as possible, as an effective strategy generally develops around a limited number of key ideas (Quinn, 1996).

Within a given mix of tactics, specific organizational traits such as power relationships and culture determine how best to deploy and use them. In short, the exact set of intervention mixes and their usage are contingent upon the specific characteristics of the enterprise in question (Schraaf, 2005; cf. SIA, 2005; cf. Arjoon, 2006; cf. Ayres and Braithwaite, 1992). Consequently, devising a compliance strategy is also a creative endeavor inspired by the organization's unique situation, not simply a matter of mechanically following a simple method.

National and organizational culture can be important factors in deciding on which levels to draw tactics from. An individualistic culture – emphasizing freedom, creativity and individual responsibility and autonomy – will be more likely to think in terms of individuals complying with norms (cf. Bond and Smith, 1996; Ralston et al., 1997; Goncalo and Staw, 2006). Such a culture can thus be expected to yield a strategy with relatively many tactics from the individual level, as individual employees and their individual managers will be held responsible for conforming to the (mandatory) norms. On the other hand, a collectivistic culture – placing more emphasis on organizational identity, collaboration, a shared vision, a consistent reputation and solidarity – will be more likely to focus on group conformity (ibid.). General compliance levels may be higher in such a culture, but the compliance ambitions may likewise be higher. Therefore, when the organization is faced with unsatisfactory compliance levels, a collectivistic culture can be expected to employ tactics from the collective and enterprise levels. Note that this would not make tactics at the individual level superfluous, as measures at the enterprise level (e.g. *developing guidelines for punishing* and *mandating compliance officers to provide incentives*) often imply the availability of tactics at the individual level (e.g. *imposing formal penalties* and *offering formal rewards*). In both types of culture the focus of the compliance strategy may shift as time progresses. A strategy drawing heavily on tactics from the individual level may go hand in hand with an increasing employee experience of unfairness, since colleagues may be rewarded or punished very differently for quite similar behavior (cf. Li et al., 2010). In time, the organization may find itself forced to develop enterprise-wide guidelines for punishing and rewarding.

## 8.7   Conclusions and further research

Our research constitutes a high-level and comprehensive view on the topic of internal compliance. We did not come across any publications in the organizational sciences with a similar broad scope and distanced view, acknowledging the value of all possible approaches. Most research instead focuses on the development of a compliance strategy or program from a specific perspective, with a limited set of tactics deemed superior. Such research thus presents specific instances of approaches and strategies, whereas our aim of this study was to take a more high-level perspective to provide a neutral and complete overview of generic ways to bring the organization into compliance. It is possible to analyze and position these tactics using the overall framework and typology presented in this chapter. This study has put forward rich conceptualizations of both compliance tactics and compliance strategies. Concep-tualizing internal compliance in terms of tactics and strategies also allows for a purer view on the topic than the standard view of internal controls does. Because of the bird's-eye view of this study, our conceptualizations are also presented neutrally and non-evaluative, thus without making the case for specific, superior approaches.

The core contribution of this chapter is the typology of 45 generic compliance tactics, the development of which has been based on a large interdisciplinary literature review of 134 publications drawing from both classical and state-of-the-art insights. To the best of our knowledge, such an overview has not been published before and this study thus fills a gap in the literature on internal compliance. Focusing on compliance tactics (rather than on internal controls) ensures we have captured the substantive essence instead of implementation activities. The resulting overview of tactics can be used as a reference work and inspirational resource by both practitioners and academics. A second contribution is the multidisciplinary overview of core insights presented in section 8.3, resulting in the development of the two-dimensional framework of section 8.4. The cells of this framework can be used for positioning and characterizing compliance tactics (resulting in the typology, i.e. Figure 8.2). The framework as a whole can also be used for describing and developing coherent and balanced compliance strategies. In this context, a third contribution is the broad and all-encompassing conceptualization of a compliance management strategy and the overview of crucial aspects of developing one. More in general, this chapter serves as an overview of fundamental insights into the topic of compliance, for academics and practitioners on the one hand, and for newcomers and knowledgeable professionals on the other.

Several limitations should be mentioned. As stated in section 8.4, a drawback of typologies is that their exhaustiveness cannot be guaranteed. Indeed, since we have chosen to draw from multiple disciplines, a huge number of publications on compliance could potentially be reviewed and not all of these could possibly be included in our review. As a consequence, we cannot guarantee that we have identified all existing tactics. To deal with this issue, however, we continued reviewing up to the point that the inventory of tactics was saturated, no new tactics were found, and new publications only served to support existing tactics. A second limitation, and another known drawback of typologies, is that mutual exclusiveness of the identified types cannot be guaranteed. We have dealt with this problem by mentioning explicitly when certain compliance tactics are related or overlapping. A third limitation is that the reviews were conducted by the principal researcher only. This has the advantage that it lowers the risk of inconsistent reviews. However, a drawback is that the definition of tactics will be more prone to an individual's subjectivity.

This risk was mitigated by regularly discussing the defined tactics with co-authors and other colleagues. In this context it should also be noted that the concepts of tactics and strategies have no absolute meaning in themselves and are inherently flexible, regardless of the research approach (Mintzberg, 1987).

The typology of tactics and the broad view on compliance management strategies may serve as a foundation for further research. A first opportunity for future research is putting the innovative candidate tactics to practice, for example by employing an action research approach. A second suggestion for further study is to verify whether individualistic cultures indeed mainly use individual level tactics, whereas collectivistic cultures also put enterprise and collective level tactics to use. As far as we know, this topic has not yet been studied. A third avenue for future research is to investigate more deeply how to develop a compliance management strategy, given the large amount of tactics available. For example, the elementary tactics in the typology can be used as input for studying opportunities for new and innovative strategic intervention mixes. In the process, more tactics may also be discovered or devised, not only because new ideas may drive innovation, but also because of the fact that certain cells of the framework are not filled as abundantly as others. A final area for further study is the development of additional theoretical perspectives on compliance strategies, using Mintzberg's five views or other conceptualizations.

## Appendix 8.A. Review protocol

*Research questions*
- What are the fundamental concepts in compliance?
- How can compliance tactics be classified?
- What tactics for stimulating compliance are acknowledged in the literature?
- How can compliance tactics be integrated in a compliance management strategy?

*Inclusion and exclusion criteria*
Inclusion criteria compliance tactics:
- Preferably academic (operationalized as at least one of the authors affiliated with a university at the time of writing) and peer reviewed.
- Non-peer reviewed (technical report).
- Non-peer reviewed and non-academic (practitioner publication). This is only permitted if the publication has a unique contribution to the research.
- English or Dutch (preferably English).
- The final decision on whether to review the publication is made on the basis of title, abstract and, if useful, conclusions and scanning through the publication.
- Papers in which compliance is the core topic are preferred, but this is no strict criterion.

Exclusion criteria compliance tactics:
- Publications that are neither in English nor in Dutch.

*Steps in a review of a publication*
Directly after reading the publication, two actions are undertaken.
- An entry is added to the literature database (see below).
- Relevant passages from the publication are captured in the first-codes-document (first-coding). A name (first code, representing a tactic name) is attached to it in the process. Text can be paraphrased or copy-and-pasted from the publication. These first-code texts can later be used as input for the pattern-coding process.
- If the publication contains knowledge relevant to the topic of compliance management strategy, relevant text is copied-and-pasted to the strategy section of the first-codes-document.
- If the pattern-coding process has already been started, the relevant pattern code(s) is (are) created or updated. Important criteria for defining a definitive pattern code (i.e. tactic) are mutual exclusiveness and a conceptual abstraction that allows for a clear and presentable description of the code.

*Literature database*
Directly after reviewing a publication, a relevant description of it is stored in the literature database. The following information is stored.
- ID: Number assigned automatically (primary key).
- Title: Title of the reviewed publication.
- Authors: Authors of the reviewed publication.
- Year: Year that the reviewed publication was published.
- Source: Journal, conference, etc in which the reviewed publication is published.
- Date entered: Date that this record is entered (and, in principle, that the publication is reviewed).
- Classification: Academic peer-reviewed, academic non-peer reviewed, or non-peer reviewed non-academic. If the classification is 'non-peer reviewed non-academic', then the Remarks section should explain the value of the publication.
- Discipline: Information Systems, Business, Law, Philosophy, Psychology.
- Core topics: The relevancy to one or more of our core topics. In other words, whether the publication explicitly covers:
    o Compliance as a concept (mainly for the definitions and framework)
    o Compliance tactics (for identifying tactics and filling of the framework)
    o Compliance strategy concept (for knowledge on creating strategy based on framework)
- Research question: The question or aim of the reviewed publication.
- Contribution: Free format text mentioning the contribution of the publication to our research.
- InDefinitiveSet: Whether the publication is included in the definitive set of relevant publications, and likely to be referenced in the final publication. Can also be Null (i.e. not yet decided).
- Unit of Analysis: Individual (e.g. Psychological research), Collective (teams, projects and departments within an organization), Organization (enterprise-wide), Nation (also multiple nations, i.e. an international scope).
- Quality evaluation: Subjective remarks concerning the quality of the study. Value are "Good", "Medium", "Inadequate".
- Remarks: Any other remarks, for example what significant contribution a non-peer reviewed non-academic offers (e.g. providing a unique insight or a clarifying illustration of a tactic in practice).

*Other remarks*
A publication can be relevant to more than one of our core topics.
A publication can have multiple levels of analysis, depending on the concept.
An academic book is considered to be peer-reviewed.

## Appendix 8.B. Screenshot of Literature Database

# Appendix 8.C. Coding Traceability

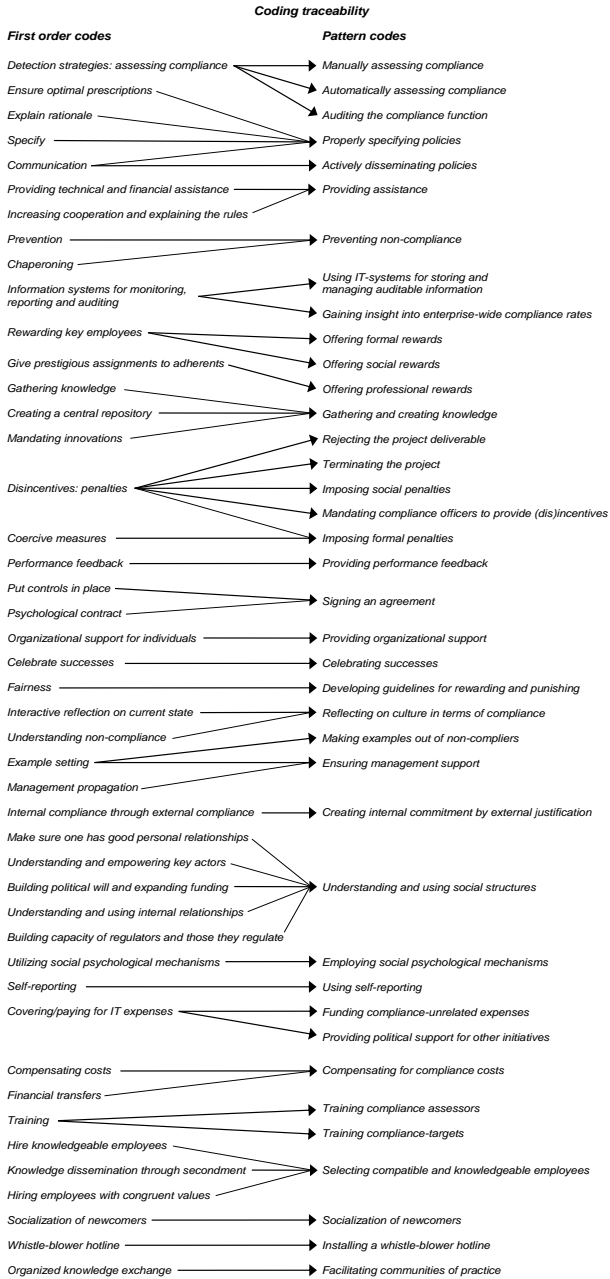

*Coding traceability*

## References

For the list of referenced literature, see the full PhD thesis:

> Foorthuis, R.M. (2012). *Project Compliance with Enterprise Architecture*. Doctoral dissertation (PhD thesis). Utrecht University, Center for Organization and Information. ISBN: 978-90-393-5834-4.

This thesis can be downloaded free of charge in PDF-format from the University Utrecht Repository at:

http://dspace.library.uu.nl/handle/1874/256014